\documentclass[a4paper,fleqn,usenatbib]{mnras}
\usepackage{newtxtext,newtxmath}

\usepackage[T1]{fontenc}
\usepackage{ae,aecompl}

\usepackage{amsmath}	
\usepackage{amssymb}	
\usepackage{graphicx}	

\usepackage{natbib}
\usepackage{longtable}
\usepackage{color}
\newcommand{\asec}{$^{\prime\prime}$}
\newcommand{\pas}{.\hskip-2pt$^{\prime\prime}$}
\def\r1415{$^{14}$N/$^{15}$N}

\def\H{N$_{2}$H$^{+}$}
\def\D{N$_{2}$D$^{+}$}
\def\15N{$^{15}$NNH$^+$}
\def\N15{N$^{15}$NH$^+$}

\def\kms{\mbox{km~s$^{-1}$}}
\def\cmc{cm$^{-3}$}
\def\cmq{cm$^{-2}$}

\def\Tex{\mbox{$T_{\rm ex}$}}

\def\kms{km\,s$^{-1}$}

\title[$^{14}$N/$^{15}$N in OMC--2 FIR4]{No nitrogen fractionation on 600 au scale in the Sun progenitor analogue OMC--2 FIR4}

\author[F. Fontani]{F. Fontani$^{1,2}$,\thanks{E-mail: fontani@arcetri.astro.it}
           G. Quaia$^{3}$,
           C. Ceccarelli$^{4}$,
           L. Colzi$^{1,3}$,
           A. L\'opez-Sepulcre$^{4,5}$,
           C. Favre$^{4}$,
           \newauthor
           C. Kahane$^{4}$,
           P. Caselli$^{2}$,
           C. Codella$^{1,4}$,
           L. Podio$^{1}$,
           S. Viti$^{6}$
         \\
$^{1}$INAF-Osservatorio Astrofisico di Arcetri, Largo E. Fermi 5, I-50125, Florence, Italy \\
$^{2}$Centre for Astrochemical Studies, Max-Planck-Institute for Extraterrestrial Physics, Giessenbachstrasse 1, 85748 Garching, Germany  \\
$^{3}$Dipartimento di Fisica e Astronomia, Universit\`a degli Studi di Firenze, I-50125 Firenze, Italy \\
$^{4}$IPAG, Universit\'e Grenoble Alpes, CNRS, F-38000 Grenoble, France \\
$^{5}$Institut de Radioastronomie Millim\'etrique (IRAM), 300 rue de la Piscine, F- 38406 Saint-Martin-dÕH\'eres, France \\
$^{6}$Department of Physics and Astronomy, UCL, Gower St., London, WC1E 6BT, UK \\
         }

\date{Accepted XXX. Received YYY; in original form ZZZ}

\pubyear{2019}

\begin{document}
\label{firstpage}
\pagerange{\pageref{firstpage}--\pageref{lastpage}}
\maketitle

\begin{abstract}
We show the first interferometric maps of the \r1415\ ratio obtained with the Atacama Large Millimeter Array 
(ALMA) towards the Solar-like forming protocluster OMC--2 FIR4. We observed \H, 
\15N, \N15\ (1--0), and \D\ (2--1), from which we derive the isotopic ratios \r1415\ and D/H. 
The target, OMC--2 FIR4, is one of the closest analogues of the environment in 
which our Sun may have formed. The ALMA images, having synthesised beam of 
$\sim 1$\pas$5 \times 1$\pas 8, i.e. $\sim 600$ au, show that the emission of the less abundant
isotopologues is distributed in several cores of $\sim 10$\asec\ (i.e. $\sim 0.02$~pc or 4000 au) 
embedded in a more extended \H\ emission. 
We have derived that the \r1415\ ratio does not vary from core to core, and our interferometric 
measurements are also consistent with single-dish observations.
We also do not find significant differences between the \r1415\ ratios computed 
from the two $^{15}$N-bearing isotopologues, \15N\ and \N15. The D/H ratio derived by comparing the 
column densities of \D\ and \H\ changes by an order of magnitude from core to core, 
decreasing from the colder to the warmer cores. Overall, our results indicate that: (1) \r1415\ 
does not change across the region at core scales, and (2) \r1415\ does not depend on 
temperature variations. 
Our findings also suggest that the $^{14}$N/$^{15}$N variations found in pristine Solar System 
objects are likely not inherited from the protocluster stage, and hence the reason has to be found 
elsewhere.
\end{abstract}


%
\section{Introduction}
\label{intro}

One of the unsolved mysteries about the Solar System is why the nitrogen
isotopic ratio, {\it R}=$^{14}$N/$^{15}$N, was $\sim$440 in the Proto
Solar Nebula (PSN, Owen et al.~\citeyear{owen01}, Fouchet et al.~\citeyear{fouchet04}, 
Marty et al.~\citeyear{marty10}), while now it is $\sim$270 in the Earth
atmosphere (Marty et al.~\citeyear{marty09}), $\sim$140 in comets (Manfroid et
al.~\citeyear{manfroid09}, Shinnaka et al.~\citeyear{shinnaka16}), and
50--300 in the Insoluble Organic Matter (IOM, e.g.~Bonal et al.~\citeyear{bonal09};
Matrajt et al.~\citeyear{matrajt12}, Nittler et al.~\citeyear{nittler18}), and
Soluble Organic Compounds (SOC, e.g.~Pizzarello \& Holmes~\citeyear{peh09}, Pizzarello
~\citeyear{pizzarello14}, Chan et al.~\citeyear{chan14}) of meteorites. What causes such 
variations was (and still is) puzzling astronomers and cosmochemists for decades.
It is now clear that there are up to three different reservoirs of nitrogen in
the Solar System, which have distinct N isotopic ratios (see for example
the discussion in F\"uri \& Marty~\citeyear{fem15}): the PSN, where {\it R}$\sim$440;
the inner Solar System, in which planets and bulk meteorites appear enriched in
$^{15}$N by a factor $\sim 1.6$ with respect to the PSN; the cometary ices, enriched 
up to a factor $\sim 3$ relative to the PSN, although the inner Solar System reservoir
could be a mixture of the two "extreme" values: the Sun and the cometary material.
Nonetheless, it remains the question of what causes the relatively large range 
of {\it R} in cometary and meteoritic material, and whether this has an ISM origin.

One popular explanation for the nitrogen isotopic fractionation has
been that, as for the hydrogen isotopic one, it has to be
attributed to (low) temperature effects (Terzieva \& Herbst~\citeyear{teh00}, 
Rodgers \& Charnley~\citeyear{rec08}, Furuya \& Aikawa~\citeyear{fea18}). 
However, several theoretical studies have excluded low temperature
isotopic exchange reactions as the main way to enhance $^{15}$N
in molecular species (e.g.~Wirstr\"{o}m et al.~\citeyear{wirstroem12},
Roueff et al.~\citeyear{roueff15}, Wirstr\"{o}m \& Charnley~\citeyear{wec18},
Loison et al.~\citeyear{loison19}). At odd with theory, some observations of cold
prestellar and protostellar objects seem to show variations of one order of magnitude
($\sim 100 - 1000$) in the N isotopic ratio, which even depends on the used molecule 
(Womack et al.~\citeyear{womack92}, Caselli \& Ceccarelli~\citeyear{cec12}, 
Bizzocchi et al.~\citeyear{bizzocchi13}, Hily-Blant et al.~\citeyear{hily-blant13},~\citeyear{hily-blant17}, 
Daniel et al.~\citeyear{daniel13}, \citeyear{daniel16}, Fontani et al.~\citeyear{fontani15a}, 
Guzman et al.~\citeyear{guzman17}, Zeng et al.~\citeyear{zeng17},
Redaelli et al.~\citeyear{redaelli18}, Colzi et al.~\citeyear{colzi18a}, De Simone et
al.~\citeyear{desimone18}). 

Kahane et al.~(\citeyear{kahane18}; see also De Simone et
al.~\citeyear{desimone18}) obtained large-scale ($\sim$10000 au)
high-precision observations of the five brightest N-bearing molecules
(HCN, HNC, CN, HC$_3$N and N$_2$H$^+$) towards one of the best known 
analogue of the environment in which the Solar System was born, OMC--2 FIR4. 
They derived the same {\it R} value, $\sim$270, in all the five molecules, 
regardless on the CN- or NH- bond. This measured value is slightly lower than 
the most recent estimate in the interstellar medium nowadays (375$\pm$60, 
Colzi et al.~\citeyear{colzi18a}),
although measurements in the local ISM show a large spread of values 
(Hily-Blant et al.~\citeyear{hily-blant17} and reference therein). 
However, these single-dish measurements provide average values over
$\sim$10000 au, namely on the whole protocluster, so that it cannot
be excluded the presence of local spots enriched (or depleted) in $^{15}$N. 
In fact, differences of a factor 2 on {\it R} can be measured from \H\
when {\it R} is measured on angular scales that can resolve the dense
cores from the extended, diffuse envelope, as found by Colzi et al.~(\citeyear{colzi19}) 
in a high-mass protocluster. 

In order to address this possibility, we obtained new ALMA
observations towards OMC--2 FIR4 of N$_2$H$^+$, $^{15}$NNH$^+$ and
N$^{15}$NH$^+$ lines with a spatial resolution of about 600 au. We
simultaneously also observed the N$_2$H$^+$ deuterated form,
N$_2$D$^+$, which provides a sort of reference for isotope
fractionation due to temperature.

This paper is organised as follows. The OMC--2 FIR4 background is reported in 
Sect.~\ref{sec:source-background}; the observations are described in Sect.~\ref{obs}; the
results are presented in Sect.~\ref{res}, and discussed in Sect.~\ref{discu}, where we 
also give the main conclusions of this paper.

\section{Source background}\label{sec:source-background}
The target of this work is the young protocluster OMC--2 FIR4 in the Orion 
star-forming region, at a distance of ~388$\pm 5$~pc (Kounkel et al.~\citeyear{kounkel17}). 
It lies in between two other young protostars: FIR3 (also known as SOF 2N or
HOPS 370, Adams et al.~\citeyear{adams12}), about 30\asec\ north-west,
and FIR5 (SOF 4 or HOPS 369, Adams et al.~\citeyear{adams12}), about
20\asec\ south-east (Mezger et al.~\citeyear{mezger90}). OMC--2 FIR4
is itself a young protocluster that harbours several embedded low- and
intermediate-mass protostars (Shimajiri et al.~\citeyear{shimajiri08},
L\'opez-Sepulcre et al.~\citeyear{lopez13}). Its unicity is due to the
fact that observations have suggested the exposition of OMC--2 FIR4 to
a dose of energetic particles very similar to that experienced by the
young Solar System. Although the source of these energetic particles is still under debate, 
and it is not clear if they are originated in the cluster itself from nascent protostars,
or from nearby external sources (Fontani et al.~\citeyear{fontani17}, Osorio et 
al.~\citeyear{osorio17}), such energetic irradiation responsible for an enhanced cosmic
ray ionisation rate was confirmed by three independent studies 
(Ceccarelli et al.~\citeyear{ceccarelli14a}; Fontani et al.~\citeyear{fontani17}; Favre et al.~\citeyear{favre18}).
This, and the increasing evidence that the Sun was born in a crowded
cluster of stars rather than in an isolated clump (Adams~\citeyear{adams10}, 
Lichtenberg et al.~\citeyear{lichtenberg19}) make OMC--2 FIR4 one of the best and
closest analogues of what must have been the environment of our Sun at
the very beginning of its formation. In this context, the study of the
N isotopic fractionation towards OMC--2 FIR4 provides constraints
on the N isotopic fractionation in an environment similar to the one
in which our Sun may have been born.

\begin{table*}
\begin{center}
\caption{Observational and spectroscopic parameters of the observed lines.}
\small
\tabcolsep 0.1cm
\label{observations}
\begin{tabular}{lcccccc}
\hline
band - molecule  & Baseline & $T_{\rm sys}$ & pwv & int. time & $\theta_{\rm SB}$ & 1$\sigma$$^{(a)}$ \\
         & range (m)   &   (K)                 & (mm) & (min) & (\asec) & (mJy) \\
         \hline
         3 - N$_2$H$^+$ & 15--491 & $50 - 130$ & $3.2$ & $136$ & $1.5$\asec$\times 1.8$\asec & $\sim 1.5$ \\
         3 - $^{15}$NNH$^+$ & 15--491 &  $50 - 100$ & $3.2$ & $136$ & $1.5$\asec$\times 1.8$\asec &  $\sim 1.2$ \\
         3 - N$^{15}$NH$^+$ & 15--491 &  $50 - 100$ & $3.2$ & $136$ & $1.5$\asec$\times 1.8$\asec &  $\sim 1.2$ \\
         4 - N$_2$D$^+$ & 15--321 & $60 - 100$ & $3.1$  & $32$ & $1.4$\asec$\times 1.8$\asec & $\sim 3$ \\
         \hline
 bandwidth         & transition  & Rest $\nu$$^{(b)}$ & $E_{\rm u}$$^{(b)}$ & $S_{\rm ij}\mu^2$$^{(b)}$ & $A_{\rm ij}$$^{(b)}$ & $\Delta V$$^{(c)}$ \\
          &                  &  (GHz)        &   (K)             &  (D$^2$)         & (s$^{-1}$)  & (\kms) \\
          \hline
         93.14 -- 93.22 &  N$_2$H$^+$ (1--0) & 93.173402  & 4.47 & 104  & 3.6$\times 10^{-5}$ & $\sim 0.2$ \\
         90.21 -- 90.32 & $^{15}$NNH$^{+}$ (1--0) & 90.263833 & 4.33 & 35  & 3.3$\times 10^{-5}$ & $\sim 0.4$ \\ 
         91.15 -- 21.26 & N$^{15}$NH$^{+}$ (1--0) & 91.205695 & 4.47 & 35  & 3.4$\times 10^{-5}$ & $\sim 0.4$ \\
         154.15 -- 154.38  &   N$_2$D$^+$ (2--1) & 154.217011  & 11.1 & 208  &  5.9$\times 10^{-4}$ & $\sim 0.2$\\
         \hline
\end{tabular}
\end{center}
$^{(a)}$ root mean square (rms) noise per channel in each spectral window; \\
$^{(b)}$ taken from the Cologne Molecular Database for Spectroscopy 
(CDMS; Endres et al.~\citeyear{endres}); \\
$^{(c)}$ spectral resolution. \\
\end{table*}

\section{Observations}
\label{obs}

Observations towards OMC--2 FIR4 using 40 antennas of the Atacama Large Millimeter Array (ALMA) 
in Cycle 4 were carried out as part of the project 
2016.1.00681.S (PI: F. Fontani), in Band 3 (3~mm) in Dec. 23 -- 25, 2016, and in band 4 (2~mm) in Mar. 11, 2017.
The correlator was configured in four different spectral windows to cover lines of \H, \15N, \N15 (1--0) at 
3~mm, and \D\ (2--1) at 2~mm. Relevant spectral parameters are given in Table~\ref{observations}. 
Flux and bandpass calibration were obtained through observations of J0423-0120. Visibility phases and amplitudes 
were calibrated on quasar J0541-0541. Some important observational parameters (baseline range, precipitable 
water vapour, system temperature, on-source total observing time, synthesised beam, and spectral resolution) are reported
in Table~\ref{observations}. The coordinates of the phase centre were RA=05$^{\rm h}35^{\rm m}27$\pas 0, 
Dec=--05$^{\circ}09^{\prime}56$\pas 8. 

The data were calibrated using standard ALMA calibration scripts of the Common Astronomy 
Software Applications (CASA\footnote{CASA is developed by an international consortium of scientists based at the National Radio Astronomical Observatory (NRAO), the European Southern Observatory (ESO), the National Astronomical Observatory of Japan (NAOJ), the Academia Sinica Institute of Astronomy and Astrophysics (ASIAA), the CSIRO division for Astronomy and Space Science (CASS), and the Netherlands Institute for Radio Astronomy (ASTRON) under the guidance of NRAO.}, version 4.7.0) package. 
The calibrated data cubes were converted in fits format and analysed in GILDAS\footnote{https://www.iram.fr/IRAMFR/GILDAS/}
format, and then imaged and deconvolved with software MAPPING of the GILDAS package using standard 
procedures. Continuum subtraction was performed by taking the line-free channels around the lines in each
individual spectral window, and subtracted from the data directly in the {\it (u,v)}-domain. 
The nominal maximum recoverable scale (MRS) was $\sim 25$\asec\ in Band 3 and $\sim 19$\asec\ in Band 4.

\section{Results}
\label{res}

\begin{table*}
\begin{center}
\caption{Peak flux densities, $F_{\nu}$, of the faintest hyperfine component in the spectra
shown in Fig.~\ref{spectra-tot} (F$_1$=0--1 of \H, and F=0--1 of \15N\ and \N15\ (1--0)), and 
their ratios, {\it R}, calculated from both \15N\ and \N15. 
$F_{\nu}$ has been estimated by fitting the lines with MADCUBA (Sect.~\ref{res-1415}). Their
error bars include the calibration uncertainty on the absolute flux density scale of 10$\%$,
and the 1$\sigma$ rms in the spectrum. This latter was computed, to be conservative, in each
region from the line-free channels around each detected transition. 
In the uncertainties on {\it R}, calculated from the propagation of the errors, the calibration 
errors on $F_{\nu}$ cancel out because the compared spectra were calibrated in the same
data cube.}
\begin{tabular}{c c c c c c c }
\hline
     Region    & \multicolumn{3}{c}{$F_{\nu}$}	& &  \multicolumn{2}{c}{{\it R}}   \\
                        \cline{2-4} \cline{6-7}
			& (Jy)	& (mJy)	& (mJy)	&	&                     &               \\
			&    N$_2$H$^+$ (1--0)   &   $^{15}$NNH$^+$ (1--0)    & N$^{15}$NH$^+$ (1--0)  &  &  ${\rm ^{15}NNH^+}$ &   ${\rm N^{15}NH^+}$   \\
			&    F$_1$=0--1          &      F=0--1                      &   F=0--1                      &  &                                     &                                       \\                            
\hline
FIR4-tot		&11$\pm$1	&	40$\pm$10 & 40$\pm$10 &	& 280$\pm$60	& 280$\pm$50  \\
FIR4-east		&1.8$\pm$0.2	&	5$\pm$2	& 9$\pm$3  &   & 360$\pm$140	 &200$\pm$70  \\
FIR4-west		&10$\pm$1	&	30$\pm$10 & 30$\pm$10  &  &330$\pm$80   &330$\pm$80  \\
FIR4-peak		&4.4$\pm$0.5	&	12$\pm$6	& 12$\pm$6  &  &370$\pm$150  &370$\pm$150 \\
FIR4-north        &2.7$\pm$0.3   &      8$\pm$4   & 10$\pm$5  &  &330$\pm$120  &270$\pm$100  \\
\hline
\end{tabular}
\label{table:table1415}
\end{center}
\end{table*}%

\begin{table*}
\begin{center}
\caption{Best fit peak velocities ($V_{\rm p}$) and full widths at half maximum (FWHM)
of the $^{15}$N-bearing lines obtained by fitting the hyperfine structure of the lines shown in Fig.~\ref{spectra-tot}
with MADCUBA (Sect.~\ref{res-1415}).}
\begin{tabular}{c c c c c c c c }
\hline
core   & \multicolumn{3}{c}{$^{15}$NNH$^+$} & & \multicolumn{3}{c}{N$^{15}$NH$^+$} \\
\cline{2-4} \cline{6-8}
          &    $V_{\rm p}$    &   FWHM  & $\tau_{\rm F=0-1}$$^{(a)}$  &  &    $V_{\rm p}$    &   FWHM  & $\tau_{\rm F=0-1}$$^{(a)}$ \\
          &     \kms\              &    \kms\    &                      & &     \kms\              &    \kms\   &                \\      
\hline
FIR4-tot  & 11.3$\pm$0.2 & 1.0$\pm$0.2 & 0.005$\pm$0.002 &  & 11.4$\pm$0.2 & 1.1$\pm$0.2  & 0.005$\pm$0.002 \\ 
FIR4-east & 11.4$\pm$0.2 & 0.9$\pm$0.2 & 0.001$\pm$0.0002 & & 11.5$\pm$0.2 & 0.8$\pm$0.2 & 0.001$\pm$0.0002 \\ 
FIR4-west & 11.3$\pm$0.2 & 1.0$\pm$0.2 & 0.001$\pm$0.0002 & & 11.4$\pm$0.2 & 1.1$\pm$0.2  & 0.002$\pm$0.0005 \\ 
FIR4-peak & 11.2$\pm$0.2 & 1.3$\pm$0.2 & 0.001$\pm$0.0002 & & 11.2$\pm$0.2 & 1.5$\pm$0.2  & 0.001$\pm$0.0002 \\ 
FIR4-north & 11.3$\pm$0.2 & 0.6$\pm$0.2 & 0.003$\pm$0.001 & & 11.2$\pm$0.2 & 0.6$\pm$0.2 & 0.002$\pm$0.0005 \\ 
\hline
\end{tabular}
\label{table:fitR}
\end{center}
$^{(a)}$ optical depth of the F=0--1 hyperfine component (see Fig.~\ref{spectra-tot}) 
\end{table*}

\begin{figure*}
\centering
{\includegraphics[width=12cm,angle=0]{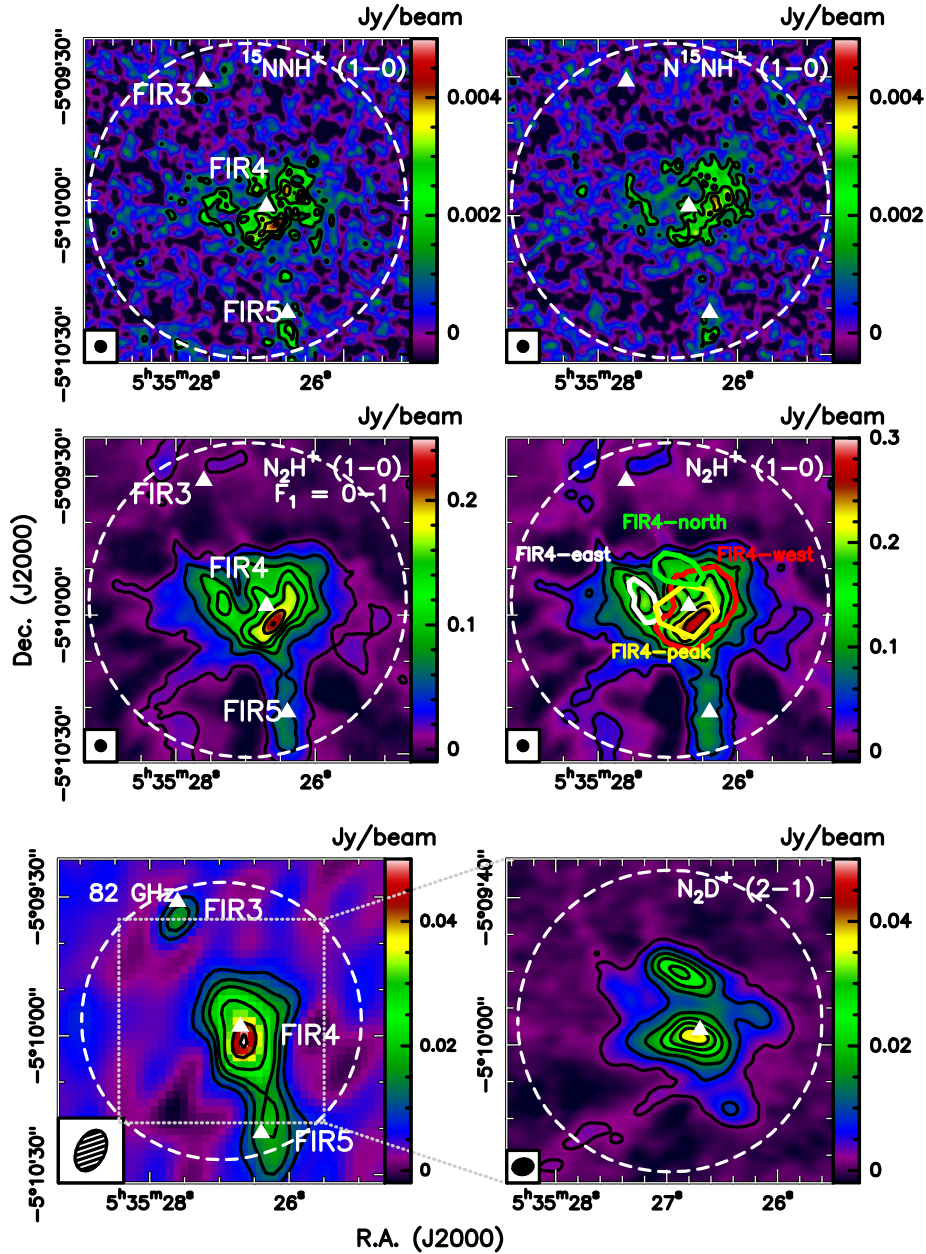}
}
\caption{Averaged emission of the \H\ isotopologues studied in this work obtained
over the whole line profile, unless when differently specified. The maps show,
pixel per pixel, the arithmetic mean of the flux density calculated over the line profiles,
and are equivalent to the integrated intensity maps, used more often in the literature, 
when multiplied by the velocity interval (given below) over which the intensity is averaged.
{\it Top panels:} (from left to right) \15N(1--0) emission averaged over the velocity
range 8.18 -- 13.72~\kms\ (contours start from the 3$\sigma$ rms level of the
averaged map, 1.5 mJy/beam, and are in steps of 1 mJy/beam), and the \N15(1--0) 
emission averaged over the velocity range 0.29 -- 19.30~\kms\ (first contour and 
step are the same as in the \15N\ map. Both velocity ranges include all hyperfine 
components. Please note that the noise of the average map is equivalent to that 
of a spectrum smoothed to a spectral resolution equal to the velocity interval covered 
by the considered channels (hence, much smaller than that of the spectral cube with
resolution $\sim 0.4$~\kms, listed in Table~\ref{observations}.
\newline
{\it Middle panels:} (from left to right) emission on the hyperfine component 
\H\ (1--0) F$_1 = 0-1$, averaged over the velocity interval 0.38 -- 3.72~\kms\ 
(contours start from 12 mJy/beam, corresponding to the 3$\sigma$ rms level of
the average map, and are in steps of 36 mJy/beam), and total \H\ (1--0) 
emission averaged over all hyperfine components in the velocity interval 0.38 -- 18.48~\kms. 
This plot also shows the four contours in which we have extracted the spectra that have been 
used to derive and discuss the \r1415\ and D/H ratios: "FIR4-west" (in red), "FIR4-east" (in white), 
"FIR4-peak" (in yellow), "FIR4-north" (in green). Spectra were also extracted from a region 
called "FIR4-tot", which is not shown because it is the union of polygons "FIR4-west", 
"FIR4-east", and "FIR4-peak". 
\newline
{\it Bottom panels:} (from left to right) 82~GHz continuum observed with NOEMA
(Fontani et al.~\citeyear{fontani17}), and \D(2--1) averaged over the velocity range 
2.87 -- 16.79~\kms, which includes all the hyperfine components (contours start from 
2.1 mJy/beam, corresponding to the 3$\sigma$rms level of the average map,
and are in steps of 6.3 mJy/beam). The map shown in the bottom right panel
is an enlargement of the region identified by the dashed square in the bottom left panel.
\newline
{\it In each frame:} the ellipse in the bottom left 
corner shows the synthesised beam (see Table~\ref{observations} for the ALMA maps, and
Fontani et al.~\citeyear{fontani17} for the NOEMA continuum map), while the dashed circle 
depicts the ALMA primary beam ($\sim 68$\asec\ and $\sim 41$\asec\ in band 3 and 4, 
respectively). 
The wedge on the right indicates the range of flux density (Jy/beam). The white 
triangles indicate the position of the far-infrared sources FIR3, FIR4 and FIR5 
(see Sect.~\ref{intro}).}
\label{map-tot}
\end{figure*}

The \H, \15N, and \N15\ (1--0) lines, and the \D\ (2--1) line, were all clearly detected 
towards OMC--2 FIR4. The maps of their intensity averaged over the full line profiles 
are shown in Fig.~\ref{map-tot}. As reference, in the same plot we show the 82~GHz 
continuum map published by Fontani et al.~(\citeyear{fontani17}). We do not show 
the ALMA continuum maps obtained from the dataset presented in Sect.~\ref{obs} because 
a high-angular resolution map of the continuum is not crucial for the analysis we make 
in this work, and a study totally devoted to the continuum emission will be presented in a 
forthcoming paper (Neri et al. in prep.).

We detect significant emission over an angular region as extended as $\sim 30$\asec\ in 
\H, and up to $\sim 20$\asec\ in \D. In particular, \H\ shows two main intensity peaks separate 
by $\sim 15$\asec\ in the east-west direction, embedded in an irregular diffuse envelope, 
while \D\ is concentrated in two cores partly overlapping along a north-south direction, whose peaks
are separated by $\sim 10$\asec. The extension of the \H\ and \D\ maps overall overlap well 
with that of the mm continuum. The rough angular size of each core is smaller than the nominal 
MRS (see Sect.~\ref{obs}) of its observing Band. The most intense \D\ core coincides 
with the strongest \H\ emission peak, while the second one is offset by $\sim 10$\asec\
to the north of the main one, detected also in \H\ but less prominent. 
The \H\ emission is more intense towards the western 
and southern portion of the protocluster, and a slight asymmetry with respect to the 
continuum can be noticed, as also found by Colzi et al.~(\citeyear{colzi19}) 
in the first (and unique so far) interferometric study of {\it R} from \H\ towards
the high-mass protocluster IRAS 05358+3543. A filamentary \H\ feature connects FIR4 
to FIR5 towards the south. This structure is present also in the continuum emission.

\begin{figure}
\centering
{\includegraphics[width=8cm,angle=0]{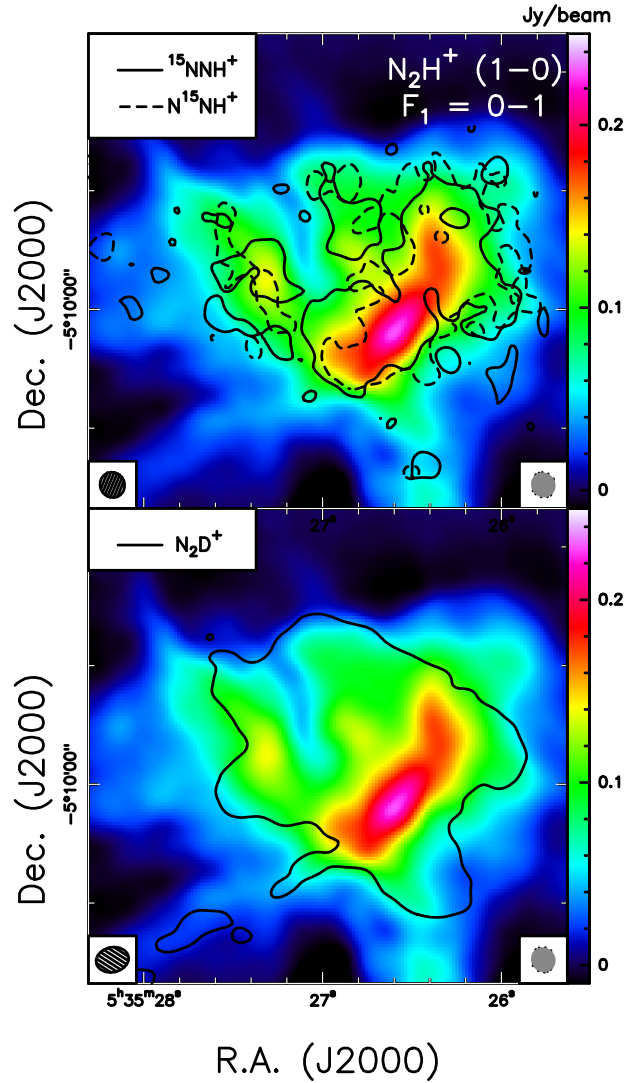}
}
\caption{Top panel: solid and dashed contours show the $3\sigma$~rms level
of the average maps of \15N\ and \N15\ (1--0, Fig.~\ref{map-tot}), respectively, 
superimposed on the average emission map of the F$_1=0-1$ hyperfine component
of \H\ (colour-scale). The ellipses in the bottom-left and bottom-right correspond to the
synthesised beams of the \15N\ and \H\ images, respectively. 
\newline
Bottom panel: same as top panel, but the depicted solid contour corresponds to the
$3\sigma$~rms level of the average map of \D\ (2--1, Fig.~\ref{map-tot}).
}
\label{fig:overlaps}
\end{figure}

The $^{15}$N isotopologues show emission much more compact than that of \H.
This could just be the consequence of the (almost) uniform sensitivity we achieved
in both the main and rare isotopologues, which likely prevents the detection of the fainter
rare isotopologues in the more diffuse envelope. Overall, the two $^{15}$N-isotopologues 
show a very similar morphology. They both mostly arise from the western 
portion of FIR4. Significant emission ($\geq 3\sigma$~rms) is also detected towards FIR5. 
Because the nature of FIR5 is still unclear, and it is not the target of the current work, this 
source will be discussed in a forthcoming paper presenting more extensively the ALMA dataset. 
To better compare the emission morphology of all \H\ isotopologues, in Fig.~\ref{fig:overlaps} 
we superimpose the average map of the emission integrated over the profile of the F=0--1 
hyperfine component of \H: even though the overall morphology is similar, clearly the rare 
isotopologues are not detected towards the diffuse \H\ emission. 

We have checked if we miss some extended flux by 
extracting spectra from a circular region corresponding to the single-dish beam of the
observations presented in Kahane et al.~(\citeyear{kahane18}): we
recover the whole flux in the $^{15}$N isotopologues, and we miss at
most $10\%$ of extended flux in \H\ (1--0), which is comparable to the
uncertainty on the flux calibration.  Hence, we can conclude that our
analysis is not affected by any significant extended emission resolved
out. The fact that there is not extended emission in \H\ in OMC--2 FIR4
makes it peculiar with respect to similar clustered star-forming regions
(e.g.~Henshaw et al.~\citeyear{henshaw14}),
in which very often interferometric \H\ emission maps suffer from extended flux
resolved out. We speculate that this could be due to a very efficient distruction 
of \H\ in the external layers, perhaps due to the high irradiation by cosmic rays, 
known to affect the chemistry of the envelope of the OMC--2 FIR4 from
different observational evidence (Ceccarelli et al.~\citeyear{ceccarelli14a}, 
Fontani et al.~\citeyear{fontani17}, Favre et al.~\citeyear{favre18}).

The complexity of the emission morphology in all lines makes it difficult to 
divide OMC--2 FIR4 in well serapated structures. Therefore, we have manually identified
five "coarse" regions defined in a very schematic way: their borders contain one 
dominant intensity peak of one isotopologue, and follow as much as possible the 
contours of one of the average maps of the rare isotopologues shown in Figs.~\ref{map-tot} 
and~\ref{fig:overlaps}. More specifically:
\begin{itemize}
\item "FIR4-east" and "FIR4-west" contain the two intensity peaks resolved in \H, 
and the borders follow roughly the $3\sigma$ rms contour of the average \15N\ and \N15\ (1--0) 
emission maps; 
\item "FIR4-peak" includes the most intense intensity peak seen in \D, and roughly follows its
$15\sigma$ rms contour to well separate this peak to the secondary one;
\item "FIR4-north" includes the less intense \D\ emission peak located $\sim 10$\asec\ north of the 
main one, and roughly follows its $15\sigma$ rms contour;
\item "FIR4-tot" coincides with the union of regions "FIR4-east", "FIR4-west" and "FIR4-peak", 
and encompasses roughly the bulk of both \15N\ and \N15\ emission. 
\end{itemize}

The spectra of \H, \15N\ and \N15\ (1--0) in flux density unit, and the \D\ (2--1) spectrum
in brightness temperature unit, extracted from these regions are shown in 
Fig.~\ref{spectra-tot}. The conversion between flux density units and brightness
temperature units for \D\ has been performed according to the equation:
$T_{\rm SB}=1.222\times10^3 F_{\nu}/(\nu^2\theta_{\rm s}^2)$,
where $\nu$ is the observing frequency in GHz units and $\theta_{\rm s}$ is the angular
source size, in arcsecond units, defined as the diameter of the circular region having the
same area of the core considered.
We display different y-axis units for the different isotopologues because of the different methods used 
to derive {\it R} and D/H, which will be discussed in the following sections. 

\begin{table*}
\begin{center}
\caption{Best fit line parameters of \H\ (1--0) and \D\ (2--1): in Cols.~2--5, we list excitation temperatures (\Tex), 
peak velocities ($V_{\rm p}$), full widths at half maximum (FWHM), and opacity of the F$_1=0-1$ hyperfine component of 
\H\ (1--0) obtained by fitting the line hyperfine structure. The fit procedure is explained in Sect.~\ref{res-1415}. 
In Cols.~6--8, we show the best fit $V_{\rm p}$, FWHM, and opacity of the main hyperfine component, $\tau_{\rm main}$, 
of \D\ (2--1), obtained fixing \Tex\ to the value given in Col.~2. This was necessary, because for \D\ the fit leaving \Tex\ as 
free parameter could not converge.
The best fit column densities are reported in Table~\ref{table:tableDH}.
}
\begin{tabular}{c c c c c c c c c}
\hline
core   & \multicolumn{4}{c}{N$_2$H$^+$} & & \multicolumn{3}{c}{N$_2$D$^+$} \\
\cline{2-5} \cline{7-9}
          & \Tex\      &    $V_{\rm p}$    &   FWHM  & $\tau_{\rm F_1=0-1}$  & &    $V_{\rm p}$    &   FWHM  & $\tau_{\rm main}$$^{(a)}$ \\
          &   K         &     \kms\              &    \kms\   &                                         & &     \kms\              &    \kms\   &  \\      
\hline
FIR4-tot  & 12.5$\pm$0.6 & 11.3$\pm$0.2 & 1.0$\pm$0.2 & 0.4$\pm$0.1 &  & 10.6$\pm$0.1& 1.0$\pm$0.1  & 0.012$\pm$0.001\\
FIR4-east & 14.6$\pm$0.5 & 11.4$\pm$0.2 & 0.8$\pm$0.2 & 0.35$\pm$0.07 & & 11.0$\pm$0.1 & 0.4$\pm$0.1 & 0.014$\pm$0.001 \\  
FIR4-west & 12.3$\pm$0.6 & 11.3$\pm$0.2 & 1.0$\pm$0.2 & 0.40$\pm$0.08 & & 10.8$\pm$0.1 & 1.0$\pm$0.1 & 0.014$\pm$0.001 \\
FIR4-peak & 12.9$\pm$0.9 & 11.2$\pm$0.2 & 1.3$\pm$0.2 & 0.30$\pm$0.1 & & 10.6$\pm$0.1 & 0.7$\pm$0.1 & 0.032$\pm$0.004 \\
FIR4-north & 11.3$\pm$0.4 & 11.2$\pm$0.2 & 0.5$\pm$0.2 & 0.35$\pm$0.08 & & 11.2$\pm$0.2 & 0.4$\pm$0.2  & 0.068$\pm$0.005 \\
\hline
\end{tabular}
\label{table:fitD}
\end{center}
$^{(a)}$ optical depth of the main hyperfine component with quantum numbers F$_1=3-2$, F=4--3 (e.g.~Gerin et al.~\citeyear{gerin01}), derived with MADCUBA (Sect.~\ref{res-DH});
\end{table*}

\begin{table*}
\begin{center}
\caption{Same as Table~\ref{table:fitD}, fixing \Tex\ to 35~K.
The best fit column densities are reported in Table~\ref{table:tableDH}.
}
\begin{tabular}{c c c c c c c c c}
\hline
core   & \multicolumn{4}{c}{N$_2$H$^+$} & & \multicolumn{3}{c}{N$_2$D$^+$} \\
\cline{2-5} \cline{7-9}
          & \Tex\      &    $V_{\rm p}$    &   FWHM  & $\tau_{\rm F_1=0-1}$  & &    $V_{\rm p}$    &   FWHM  & $\tau_{\rm main}$$^{(a)}$ \\
          &   K         &     \kms\              &    \kms\   &                                         & &     \kms\              &    \kms\   &  \\      
\hline
FIR4-tot  & 35 & 11.3$\pm$0.2 & 1.2$\pm$0.2 & 0.12$\pm$0.01 &  & 10.6$\pm$0.1& 1.1$\pm$0.1  & 0.003$\pm$0.001\\
FIR4-east & 35 & 11.4$\pm$0.2 & 0.9$\pm$0.2 & 0.15$\pm$0.01 & & 11.0$\pm$0.1 & 0.4$\pm$0.1 & 0.005$\pm$0.001 \\  
FIR4-west & 35 & 11.3$\pm$0.2 & 1.3$\pm$0.2 & 0.11$\pm$0.01 & & 10.8$\pm$0.1 & 1.0$\pm$0.1 & 0.004$\pm$0.001 \\
FIR4-peak & 35 & 11.2$\pm$0.2 & 1.4$\pm$0.2 & 0.10$\pm$0.01 & & 10.6$\pm$0.1 & 0.7$\pm$0.1 & 0.009$\pm$0.004 \\
FIR4-north & 35 & 11.2$\pm$0.2 & 0.6$\pm$0.2 & 0.15$\pm$0.01 & & 11.2$\pm$0.2 & 0.4$\pm$0.2  & 0.015$\pm$0.005 \\
\hline
\end{tabular}
\label{table:fitD2}
\end{center}
$^{(a)}$ optical depth of the main hyperfine component with quantum numbers F$_1=3-2$, F=4--3 (e.g.~Gerin et al.~\citeyear{gerin01}), derived with MADCUBA (Sect.~\ref{res-DH});
\end{table*}

\begin{table*}
\begin{center}
\caption{Same as Table~\ref{table:fitD}, fixing \Tex\ to 45~K.
The best fit column densities are reported in Table~\ref{table:tableDH}.
}
\begin{tabular}{c c c c c c c c c}
\hline
core   & \multicolumn{4}{c}{N$_2$H$^+$} & & \multicolumn{3}{c}{N$_2$D$^+$} \\
\cline{2-5} \cline{7-9}
          & \Tex\      &    $V_{\rm p}$    &   FWHM  & $\tau_{\rm F_1=0-1}$  & &    $V_{\rm p}$    &   FWHM  & $\tau_{\rm main}$$^{(a)}$ \\
          &   K         &     \kms\              &    \kms\   &                                         & &     \kms\              &    \kms\   &  \\      
\hline
FIR4-tot  & 45 & 11.3$\pm$0.2 & 1.2$\pm$0.2 & 0.10$\pm$0.01 &  & 10.6$\pm$0.1& 1.1$\pm$0.1  & 0.002$\pm$0.0005\\
FIR4-east & 45 & 11.4$\pm$0.2 & 0.9$\pm$0.2 & 0.11$\pm$0.01 & & 11.0$\pm$0.1 & 0.4$\pm$0.1 & 0.004$\pm$0.001 \\  
FIR4-west & 45 & 11.3$\pm$0.2 & 1.3$\pm$0.2 & 0.08$\pm$0.01 & & 10.8$\pm$0.1 & 1.0$\pm$0.1 & 0.003$\pm$0.001 \\
FIR4-peak & 45 & 11.2$\pm$0.2 & 1.4$\pm$0.2 & 0.08$\pm$0.01 & & 10.6$\pm$0.1 & 0.7$\pm$0.1 & 0.007$\pm$0.002 \\
FIR4-north & 45 & 11.2$\pm$0.2 & 0.6$\pm$0.2 & 0.11$\pm$0.02 & & 11.2$\pm$0.2 & 0.4$\pm$0.2  & 0.012$\pm$0.005 \\
\hline
\end{tabular}
\label{table:fitD3}
\end{center}
$^{(a)}$ optical depth of the main hyperfine component with quantum numbers F$_1=3-2$, F=4--3 (e.g.~Gerin et al.~\citeyear{gerin01}), derived with MADCUBA (Sect.~\ref{res-DH});
\end{table*}

\subsection{$^{14}$N/$^{15}$N}
\label{res-1415}

From the \H, \15N\ and \N15\ (1--0) spectra shown in Fig.~\ref{spectra-tot}, we have
first derived the \r1415\ ratios, {\it R}, following this approach:
we have divided the flux density peak of the F$_1$=0--1 hyperfine component of \H\ (1--0), with 
rest frequency 93.17613~GHz, by that of the F=0--1 one of both \15N\ and \N15\ (1--0), at rest 
frequencies 90.2645 and 91.2086~GHz, respectively. These components are indicated in
Fig.~\ref{spectra-tot}. Full spectroscopic parameters of the hyperfine structure of the 
three lines are given, e.g., in Kahane et al.~(\citeyear{kahane18}) and Colzi et al.~(\citeyear{colzi19}). 
If the compared hyperfine components are optically thin, have the same line width, and have the 
same relative intensity (which indeed is $\sim 0.1111$ for all isotopologues for the considered hyperfine 
components, see e.g.~Dore et al.~\citeyear{dore09} for \15N\ and \N15\ (1--0), and the Jet Propulsion
Laboratory (JPL) catalog\footnote{https://spec.jpl.nasa.gov/ftp/pub/catalog/catform.html} for \H\ (1--0)), 
then the ratio between the peak flux densities of these components
is equivalent to the ratios between the total flux densities of the lines. Kahane et al.~(\citeyear{kahane18}) compute the 
\r1415\ ratio from the total line intensity, which is, however, derived by them from the velocity-integrated intensity 
of the same isolated hyperfine component considered by us, and also assuming optically thin conditions. 
Hence, if the considered hyperfine components of the different isotopologues have the same line width, 
and are optically thin, the methods are equivalent, and the ratios between the peak intensities of the considered 
hyperfine components and the total line intensities are the same.

The best fit peak fluxes, $F_{\nu}$, of the analysed hyperfine components have been obtained with
the software MADCUBA\footnote{Madrid Data Cube Analysis on ImageJ is a 
software developed in the Center of Astrobiology (Madrid, INTA-CSIC) 
to visualise and analyse single spectra and datacubes (Mart\'in et al., in prep., 
Rivilla et al.~\citeyear{rivilla16}).}, which performs a fit to the lines hyperfine structure 
creating a synthetic profile assuming for all components a single excitation temperature, 
\Tex, line width at half maximum, FWHM, and separation in velocity given by the laboratory
value. The software also computes the best fit line velocity at the intensity peak, $V_{\rm p}$,
the opacity of the various components, and the total column density of the molecule. 
The best fit $F_{\nu}$ are listed in Table~\ref{table:table1415}, where we also give {\it R} 
computed in the four regions identified in Fig.~\ref{map-tot}. 
By comparing the 1$\sigma$ rms noise level in the spectra (Table~\ref{observations}) 
with the peak intensities of the analysed hyperfine components (Table~\ref{table:table1415}),
we can see that the faintest hyperfine components are all detected with a signal-to-noise 
ratio $\geq 5$, except for \15N\ in FIR4-east for which the signal-to-noise ratio is $\sim 4.2$.

The fits shown in Fig.~\ref{spectra-tot} provide generally good results, except in some \N15\ spectra 
in which an extra feature at $\sim 9$~\kms\ above $3\sigma$ in the residuals is revealed. 
Given that at the feature frequency ($\sim 91206.5$~GHz) there are no lines of other species that 
can reasonably be attributed to this feature, it could be a second velocity component. However, this 
feature is significantly detected only in the main hyperfine component of \N15, which is not the 
one we use in our analysis. Close to the component used in our analysis we never detect a significant
secondary peak. Therefore, this extra feature has probably no influence on the isotopic ratios that 
we derive.

Fig.~\ref{ratios-15n} shows the comparison between {\it R} derived in the different regions
shown in Fig.~\ref{map-tot}, and that obtained by Kahane et al.~(\citeyear{kahane18}) with
the IRAM-30m telescope:
it is apparent that in OMC--2 FIR4 {\it R} does not change neither from region to region
nor going from the single-dish scale to the interferometric scale, within the errobars. 
The isotopic ratios derived in each core from the two isotopologues do not show significant 
differences between them as well. The core in which the two values show the largest 
discrepancy is "FIR4--east" (360$\pm 140$ and 200$\pm 70$ from \15N\ and \N15, respectively), 
but even here the two estimates are consistent within the (large) error bars.

As pointed out above, our method is based on the fact that the relative intensity of the aforementioned 
components with respect to the others of their hyperfine pattern is the same in all isotopologues
(see e.g.~Dore et al.~\citeyear{dore09}. Thus, their ratio depends only on the 
isotopic ratio \r1415\ provided that the compared hyperfine components: (1) have the same excitation
temperature, (2) have the same line widths, (3) are all optically thin. 

Condition (1) is very likely because the three transitions have very similar critical densities.
However, let us discuss better this approximation: from a non-LTE analysis, Hily-Blant et al.~(\citeyear{hily-blant13}) 
found differences in \Tex\ for lines with the same quantum numbers of the different isotopologues of HCN. 
But these differences are in all (but one) cases below $\sim 10\%$, indicating that a significantly different
\Tex\ for lines with the same quantum numbers is unlikely for isotopologues of the same species. 
Regarding the possibility that different hyperfine 
components of the same isotopologue line can have a different \Tex, Daniel et al.~(\citeyear{daniel06}) showed that 
high optical depths in \H\ (1--0) could indeed cause deviations from the line profile expected when
each component has the same excitation temperature. 
According to Table~\ref{table:table1415}, the optical depth of the F$_1=0-1$ component of 
\H\ (1--0) is in between $\sim 0.3$ and $\sim 0.4$, which translates into high total opacities of
the lines and hence possible hyperfine "anomalies" in their profiles. 
However, both theoretical (Daniel et al.~\citeyear{daniel06}) and observational (Caselli et al.~\citeyear{caselli95}) 
works show that \Tex\ of the component analysed in our work
would deviate from the local thermodynamic equilibrium value by 10-15$\%$ at most, 
and only at H$_2$ volume densities below $10^{5}$~\cmc\ (see Fig.~6 in Daniel et al.~\citeyear{daniel06}).
Because in OMC--2 FIR4 the average H$_2$ volume density is $1.2\times 10^6$~\cmc\ 
(Ceccarelli et al.~\citeyear{ceccarelli14a}), where the predicted deviations from the equilibrium \Tex\
is negligible (Fig.~6 in Daniel et al.~\citeyear{daniel06}), we are confident that hyperfine anomalies 
are not affecting significantly the \Tex\ of the analysed component. 
This is also confirmed by the qualitative excellent agreement between data 
and fits (which indeed assume a single \Tex\ for all hyperfine components) for \H\ (1--0) in 
Fig.~\ref{spectra-tot} around the F$_1$=0--1 components.

About assumptions (2) and (3), let us first discuss the \15N\ and \N15\ (1--0) lines.
As stated above, the fit performed with MADCUBA provides several parameters, among which 
$V_{\rm p}$, FWHM, and the opacity of each hyperfine component. The tool MADCUBA-AUTOFIT 
provides the best fit parameters via a non-linear least squared fit algorithm (see also 
Colzi et al.~\citeyear{colzi19}). The results of these fits are shown in Table~\ref{table:fitR} and
indicate that the FWHM of the two isotopologues are the same within the uncertainties, 
and that the optical depth of the compared components is well below 0.1. The fact that
the \15N\ and \N15\ opacities are so low in the less intense component is consistent 
with our expectations, given the low abundance of these two isotopologues. However,
we stress that in some cases the uncertainties on the opacities are underestimated. In fact,
as discussed in Mart\'in et al.~(\citeyear{martin19}), when one of the fit parameters in MADCUBA 
is fixed, then the error associated is zero, and hence the uncertainties of all quantities calculated 
from this parameter do not include its error. As we will illustrate in the following, 
sometimes we fixed \Tex, and hence in these cases the errors on the optical depth will 
be underestimated. However, the main point of this analysis is simply to confirm that the 
lines are optically thin, which indeed is confirmed.

For \H\ (1--0), for which some hyperfine components are overlapping and the line optical
depth could be higher, we have performed a more accurate analysis of the line profile
assuming three different \Tex: the best fit \Tex\ when leaving
this parameter free, and the two extreme values of kinetic temperature measured in the envelope 
of OMC--2 FIR4, namely 35 and 45~K, in previous 
works (Ceccarelli et al.~\citeyear{ceccarelli14a}, Friesen \& Pineda et al.~\citeyear{friesen17}).
The results are shown in Tables~\ref{table:fitD}, \ref{table:fitD2}, and \ref{table:fitD3}.
By comparing the parameters shown in these Tables and those reported in Table~\ref{table:fitR}, 
for each single core, we find that the \H, \15N, and \N15\ lines have the same FWHM, 
within the uncertainties, and that the optical depth of the F$_1=0-1$ component is at most 
0.4. Thus, conditions (2) and (3) are also satisfied.

For completeness, we have quantified the error that we make in
the most unfavourable case in our simplified approach: because
at most, $\tau$ of the F$_1$=0--1 component is 0.4$\pm$0.1 for FIR4-tot, and 
0.40$\pm$0.08 for FIR4-west, in both cases the peak brightness temperature should be 
corrected by the factor $\tau/(1-\exp{(-\tau)})\sim 1.21$. 
For FIR4-tot, the \r1415\ ratio would change from $280\pm 60$ and $280\pm 50$ for
\15N\ and \N15\, respectively, to about $340\pm 100$ for both, and for FIR4-west it 
would change from $330\pm 80$ from both \15N\ and \N15, to $400\pm 120$. 
These values are still consistent, within the uncertainites, to those derived in the other regions. 
Therefore, even considering the correction for the optical depth, 
the \r1415\ ratio does not change from region to region within the uncertainties.

\subsection{D/H}
\label{res-DH}

\begin{figure*}
\centering
{\includegraphics[width=17cm,angle=0]{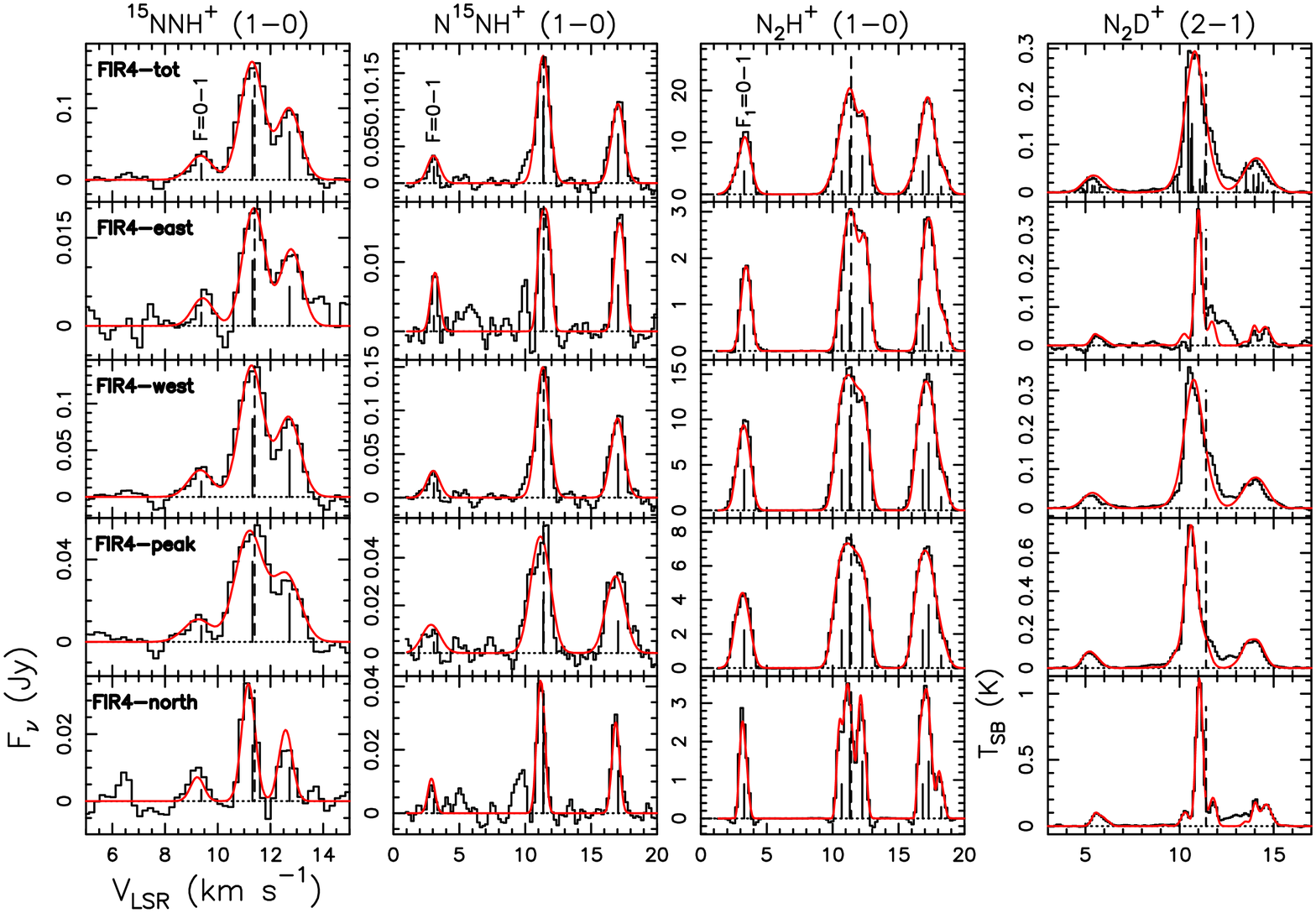}}
\caption{Spectra of, from left to right, \15N (1--0), \N15\ (1--0), \H\ (1--0) and \D\ (2--1) extracted from the five 
regions defined in Fig.~\ref{map-tot}, namely (from top to bottom): "FIR4-tot", "FIR4-east", "FIR4-west", "FIR4-peak", 
and "FIR4-north" (see Sect.~\ref{res} for details). In each spectrum,
the vertical dashes line indicates the systemic velocity of 11.4~\kms, and the horizontal dotted line
corresponds to y=0. The \15N, \N15, and \H\ spectra are in flux density units, because {\it R} is derived
from the peak fluxes of the hyperfine components labelled in the top spectra of each column. 
For the \15N, \N15, and \H\ spectra, the red curve is the best fit to the line hyperfine 
structure obtained with MADCUBA, whose components are indicated by vertical, solid lines under 
each spectrum, the length of which is proportional to the expected relative intensity in LTE;
The \D\ spectra are in brightness temperature ($T_{\rm SB}$) units and the red curve 
shows the best fit obtained with MADCUBA because the parameters reported in Sect.~\ref{res} and 
Table~\ref{table:fitD} were derived from the \D\ spectra converted in these units (see Sect.~\ref{res}). 
The hyperfine structure of \D\ (2--1) is shown only in the spectrum of FIR4-tot for clarity 
of the figure.}
\label{spectra-tot}
\end{figure*}

\begin{figure}
\centering
\includegraphics[width=8.5cm,angle=0]{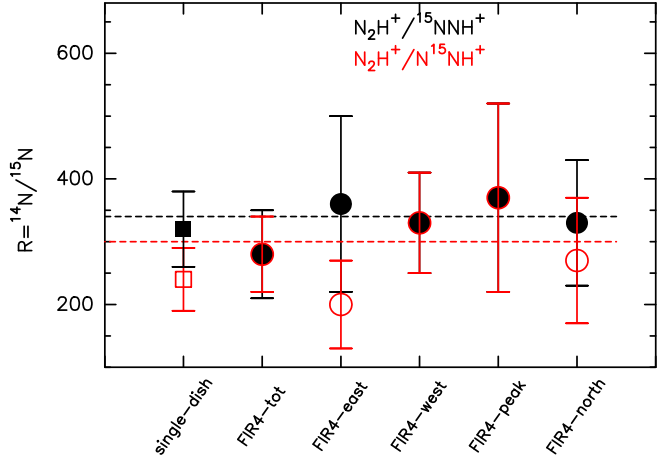}
\caption{Isotopic ratio {\it R}=\r1415\ computed 
from either \15N\ (black symbols) or \N15\ (red symbols). The ratios have been derived for the five regions 
discussed in Sect.~\ref{res} and Fig.~\ref{map-tot} following the method described in Sect.~\ref{res-1415}. 
The horizontal dashed lines correspond to the average of {\it R} in the five regions from \15N\ (black)
and \N15\ (red). 
For comparison, we also show the single-dish measurements obtained by Kahane et al.~(\citeyear{kahane18},
squares). 
It is apparent that the \r1415\ ratios do not change from region to region and are all comparable to the 
corresponding single-dish values, within the error bars. 
No statistically significant differences are found between the two $^{15}$N isotopologues either.}
\label{ratios-15n}
\end{figure}

We have derived the D/H ratio from the \H\ (1--0) and \D\ (2--1) lines. Due to the different quantum 
numbers of these transitions, we could not use the approach adopted to evaluate {\it R}. 
Therefore, the D/H ratio has been estimated by dividing the total column densities of \D\ and \H\ 
computed by fitting the lines with MADCUBA. We have fitted the lines with three temperatures,
as explained in Sect.~\ref{res-1415}.
The fits to the hyperfine structure have been performed to both \H\ and \D\ spectra
converted in "synthesised temperature" units. Generally the lines are well fitted, 
although in all \D\ (2--1) spectra a residual emission partly overlapping the hyperfine
pattern is apparent at $\sim 154.2165$~GHz (right panels in Fig.~\ref{spectra-tot}). This 
could be due to a contamination from the transition $(8_{4,4}-7_{4,3})$ of CH$_3$CHO at 154216.68~GHz 
($E_{\rm u}\sim 69$~K, $S_{\rm ij}\mu^2$=75.9 D$^2$). However, the large number of hyperfine 
components not contaminated by this excess emission (which, however, is always smaller than $5\%$ 
of the total line integrated intensity, i.e. smaller than the calibration error) allows us to well fit the 
hyperfine pattern in all \D\ spectra. 

The best fit $V_{\rm p}$, FWHM, and opacity of the main component of \D\ (2--1) are shown in 
Table~\ref{table:fitD}, where we also give the best fit parameters derived for \H\ (1--0) already presented in 
Sect.~\ref{res-1415}.
The best fit column densities are shown in Table~\ref{table:tableDH}, and
are in the range $\sim 0.7 - 1.7\times 10^{14}$ \cmq\ for \H, and $\sim 2.5 - 13.8\times 10^{11}$ \cmq\ for \D,
which translate into D/H values in between 2.6$\times 10^{-3}$ 
towards "FIR4-east" and 1.4$\times 10^{-2}$ towards "FIR4-north".
We stress that, even though the total column densities in each region change by a
factor $\sim 1.5$ assuming different excitation temperatures, the D/H ratios 
do not change within the uncertainties. 

\begin{table*}
\begin{center}
\caption{Total column densities of \H\ and \D\ and their ratio, D/H, derived fitting the spectra in Fig.~\ref{spectra-tot} 
with MADCUBA (see Sect.~\ref{res-DH}) 
assuming three different \Tex: the best fit value for \H\ reported in Table~\ref{table:fitD}, $T^{\rm fit}_{\rm ex}$,
and two fixed values, 35 and 45~K, corresponding to the extrema of the kinetic temperature range estimated 
for the envelope of OMC--2 FIR4 by Ceccarelli et al.~(\citeyear{ceccarelli14a}).
We note that although the total column densities change with \Tex\ by a factor $\sim 1.5$, the D/H ratios are 
equal within the uncertainties.}
\begin{tabular}{c c c c c}
\hline
     Region    & N(N$_2$H$^+$)		& N(N$_2$D$^+$)	& ${\rm \frac{D}{H}=\frac{N(N_2D^+)}{N(N_2H^+)}}$ \\
			& $\times$10$^{14}$(cm$^{-2})$	&$\times$10$^{11}$(cm$^{-2}$)  &	$\times$10$^{-3}$	\\
				\cline{2-4} 
			&  $T^{\rm fit}_{\rm ex}$$^{(a)}$ - 35~K - 45~K   & $T^{\rm fit}_{\rm ex}$ - 35~K - 45~K    & $T^{\rm fit}_{\rm ex}$ - 35~K - 45~K                    \\
\hline
FIR4-tot		&  1.02$\pm$0.05 - 1.23$\pm$0.04 - 1.48$\pm$0.04	& 4.9$\pm$0.1 - 7.2$\pm$0.2 - 8.5$\pm$0.2	& 5$\pm1$ - 6$\pm$1 - 6$\pm$1 \\
FIR4-east	         &     0.98$\pm$0.02 - 1.23$\pm$0.03	- 1.48$\pm$0.03	& 2.5$\pm$0.1 - 3.6$\pm$0.2 - 4.3$\pm$0.2	& 2.6$\pm$0.5 - 2.9$\pm$0.6 - 2.9$\pm$0.6 \\
FIR4-west		 & 1.05$\pm$0.06 - 1.29$\pm$0.04	- 1.51$\pm$0.05	& 5.4$\pm$0.1 - 7.8$\pm$0.2 - 9.3$\pm$0.2	& 5$\pm$1 - 6$\pm$1 - 6$\pm$1 \\
FIR4-peak		& 1.10$\pm$0.06 - 1.38$\pm$0.05 - 1.66$\pm$0.04	& 8.9$\pm$0.1 - 12.6$\pm$0.2 - 15.2$\pm$0.3	& 8$\pm$2 - 9$\pm$2 - 9$\pm$2	\\
FIR4-north       & 0.69$\pm$0.03 - 0.83$\pm$0.03 - 0.98$\pm$0.03     & 8.3$\pm$0.1 - 11.5$\pm$0.2 - 13.8$\pm$0.3  & 12$\pm$2 - 14$\pm$3 - 14$\pm$3 \\
\hline
\end{tabular}
\label{table:tableDH}
\end{center}
$^{(a)}$ listed in Col.~2 of Table~\ref{table:fitD};
\end{table*}   

\section{Discussion and conclusions}
\label{discu}

\begin{figure}
\centering
\includegraphics[width=8cm,angle=0]{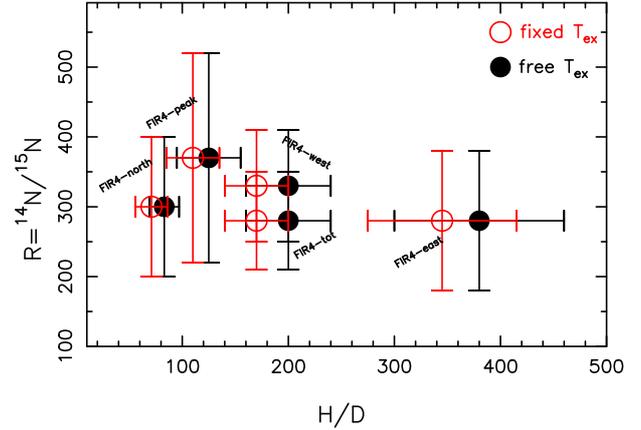}
\caption{Comparison between the isotopic ratios {\it R}=\r1415\ and H/D listed in 
Tables~\ref{table:table1415} and \ref{table:tableDH}, respectively, computed as 
explained in Sects.~\ref{res-1415} and \ref{res-DH}. 
A label identifies each of the five regions defined in Sect.~\ref{res}. 
We report the H/D ratios calculated by both fixing \Tex\ to 35--45~K (red symbols, 
the two estimates are identical within the errors) and leaving \Tex\ as a free fit parameter 
(black symbols).}
\label{ratios-d}
\end{figure}

Fig.~\ref{ratios-d} shows the H/D ratio against {\it R}: while H/D varies from $\sim 70$ in 
"FIR4-north" to $\sim 380$ in "FIR4-east", {\it R} does not. This demonstrates that there 
is no correlation between N and H fractionation, as also deduced by previous studies both
in the same molecule (Fontani et al.~\citeyear{fontani15a}) and in nitriles (Colzi et al.~\citeyear{colzi18b}). 
Finally, it is worth noticing that the highest H/D ratio measured in "FIR4-east" agrees with previous 
observations which indicate that the eastern part of the protocluster is warmer than the 
western one (Fontani et al.~\citeyear{fontani17}, Favre et al.~\citeyear{favre18}). On the
opposite, "FIR4-north", having the lowest H/D and located to the north-western part of the 
protocluster, is likely the coldest (and maybe less evolved) condensation. The D/H
ratio of \H\ is a clear evolutionary indicator in low- and high-mass dense cores, and it is well
anti-correlated to the gas temperature (Crapsi et al.~\citeyear{crapsi05}, Ceccarelli et al.~\citeyear{ceccarelli14b},
Fontani et al.~\citeyear{fontani15b}, De Simone et al.~\citeyear{desimone18}). 
Therefore, our observations also demonstrate the independence of the N fractionation 
on the gas temperature and maybe also on the core evolutionary stage, in agreement with 
the most recent theoretical predictions (see Sect.~\ref{intro}).

It is interesting and useful to make a comparison with the results obtained by Colzi et 
al. (\citeyear{colzi19}) towards the high-mass star forming region IRAS 05358+3543.
Colzi et al.~(\citeyear{colzi19}) found that {\it R} in \H\ shows an enhancement from
$\sim 100 - 220$ up to $\geq 200$ (i.e. up to a factor $\sim 2$) from the core scale of 
$\sim 5$\asec, to the diffuse emission in the envelope. 
However, even if the two works have similar angular resolution, the distance of 
IRAS 05358+3543 (1.8 kpc) allowed Colzi et al.~(\citeyear{colzi19}) to resolve
a linear scale of $\sim 0.05$ pc, or $\sim 10000$ au. At the distance of OMC--2 FIR4, this
would correspond to $\sim 25$\asec, i.e. about the total \H\ emission size in this work. 
Therefore, the two works are complementary, and our results indicate that on linear scales 
smaller than 0.05~pc, not sampled by Colzi et al.~(\citeyear{colzi19}), {\it R} remains constant. 
To investigate if {\it R} increases from envelope to protocluster scale in OMC--2 FIR4 as
found in IRAS 05358+3543, we have extracted spectra in different points of the envelope 
surrounding FIR4, in which \H\ is detected but the $^{15}$N isotopologues are not, and
hence derived lower limits for {\it R} in the envelope of FIR4.
We have found that {\it R}$\geq 100-150$. This lower limit is smaller than the values derived in 
the internal part of the protocluster, hence not sufficient to put stringent constraints and unveil 
a possible change of {\it R} from the diffuse envelope to the dense protocluster. Therefore, a change 
of {\it R} from the envelope to the inner part of the protocluster as found in IRAS 05358+3543
cannot be ruled out.

Based on our results, the $^{15}$N enrichment found in comets and protoplanetary disks
(Sect.~\ref{intro}) does not seem to be inherited from the protostellar/protocluster stage, 
even when measured at core scales. It does not even seem to vary from the pre- to the 
proto-stellar stage, because the average \r1415\ ratio measured towards OMC--2 FIR4
seems also consistent with the ratios derived in pre--stellar cores (e.g.~Daniel et al.~\citeyear{daniel13}, 
\citeyear{daniel16}). However, as stated in Sect.~\ref{intro}, care needs to be taken in this
comparison because the \r1415\ ratio in pre-stellar cores shows a huge spread of values (of 
about an order of magnitude) when considering different molecules (e.g.~Bizzocchi et 
al.~\citeyear{bizzocchi13}, Redaelli et al.~\citeyear{redaelli18}), the reasons of which 
is still not understood.

Based on our results and on the previous works in the literature, we propose hence two 
alternative scenarios for the $^{15}$N enrichment, which, however, are not able to explain 
all the observational results: 
\begin{itemize}
\item[(1)] it occurs during the protoplanetary disk stage due to, for example, selective 
photodissociation of N$_2$, as already proposed by Guzman et al.~(\citeyear{guzman17}) 
to explain the $^{15}$N enrichment in HCN (see also Visser et al.~\citeyear{visser18}). 
This, however, cannot explain the \r1415\ ratio measured in TW Hya with CN, 
similar to the pre-stellar value (Hily-Blant et al.~\citeyear{hily-blant17}); 
\item[(2)] the enrichment occurs at different stages, depending on the molecule. For example,
\H\ and HCN could be enriched at the protoplanetary disk stage.
\end{itemize}
However, limited measurements of {\it R} in protoplanetary disks and in pre- and proto-stellar objects
at core scales, have been performed so far. Therefore, to test both scenarios, comparative measurements 
of {\it R} in \H\ and other species in representatives of the different evolutionary stages of the Solar system 
are needed. 


{\it Acknowledgments.} 

We thank the anonymous Referee for his/her careful reading of the paper and 
for his/her useful comments. This work has been partially supported by the project 
PRIN-INAF 2016 The Cradle of Life - GENESIS-SKA (General Conditions in Early 
Planetary Systems for the rise of life with SKA).
CC and FF acknowledge the funding from the European Research Council (ERC) 
under the European Union's Horizon 2020 research and innovation programme, 
for the Project "The Dawn of Organic Chemistry" (DOC), grant agreement No 741002.
LC acknowledges support from the Italian Ministero del\l'Istruzione, Universit\'a e 
Ricerca through the grant Progetti Premiali 2012 - iALMA (CUP C52I13000140001).
This paper makes use of the following ALMA data: ADS/JAO.ALMA$\#$2016.0.00681.S. 
ALMA is a partnership of ESO (representing its member states), NSF (USA) and 
NINS (Japan), together with NRC (Canada), MOST and ASIAA (Taiwan), and 
KASI (Republic of Korea), in cooperation with the Republic of Chile. The Joint ALMA 
Observatory is operated by ESO, AUI/NRAO and NAOJ.
\let\oldbibliography\thebibliography
\renewcommand{\thebibliography}[1]{\oldbibliography{#1}
\setlength{\itemsep}{-1pt}} 

{}

\end{document}